\documentclass[aps,prb,twocolumn]{revtex4}
\usepackage{graphicx,amsfonts,amssymb}
\usepackage{amsmath}
\usepackage{color}

\usepackage{epstopdf}
\newcommand{\ud}[1]{{#1^{\dagger}}}

\newcommand{\ket}[1]{\left| #1\right\rangle}

\newcommand{\mean}[1]{\langle #1\rangle}

\newcommand{\abs}[1]{\left|{#1}\right|}
\newcommand{\br}[1]{\langle #1|}
\newcommand{\ke}[1]{|#1\rangle}

\newcommand{\kb}[2]{\ke{#1}\br{#2}}
\newcommand{\da}{^\dagger}
\newcommand{\pt}[1]{\left( #1 \right)}
\newcommand{\pq}[1]{\left[ #1 \right]}
\newcommand{\pg}[1]{\left\{ #1 \right\}}

\newcommand{\av}[1]{\left\langle #1 \right\rangle}
\newcommand{\al}[1]{^{(#1)}}

\newcommand{\lpq}[1]{\left[ #1 \right.}

\newcommand{\rpq}[1]{\left. #1 \right]}

\begin{document}

\title{Two-photon lasing by a single quantum dot in a high-Q microcavity}
\author{Elena del Valle,$^{1,2}$ Stefano Zippilli, $^3$ Fabrice P. Laussy, $^{2}$ Alejandro Gonzalez-Tudela,$^1$ Giovanna Morigi,$^{3,4}$ and Carlos Tejedor$^1$}

\address{$^1$ Departmento de F\'isica Te\'orica de la Materia Condensada, Universidad Aut\'onoma de Madrid, 28049 Madrid, Spain}
\address{$^2$ School of Physics and Astronomy, University of Southampton, SO171BJ, Southamtpon, UK}
\address{$^3$ Departament de F\'isica, Universitat Aut\`onoma de Barcelona, 08193 Bellaterra, Spain}
\address{$^4$ Theoretische Physik, Universit\"at des Saarlandes, D-66041 Saarbr\"ucken, Germany}

\date{\today}
\begin{abstract}
We investigate theoretically  two-photon processes in a microcavity containing one quantum dot in the strong coupling regime. The cavity mode can be tuned to resonantly drive the two-photon transition between the ground and the biexciton states, while the exciton states are far-off resonance due to the biexciton binding energy. We study the steady state of the quantum dot and cavity field in presence of a continuous incoherent pumping. We identify the regime where the system acts as two-photon emitter and discuss the feasibility and performance of realistic single quantum dot devices for two-photon lasing.
\end{abstract}
\maketitle

\section{Introduction}

Two-photon emission is an inherently nonclassical effect.  Realizing such emitters { is important} for instance in quantum telecommunication~\cite{Roadmap,ziliang}.  In this respect, an incoherently pumped source acting as two-photon laser is an attractive goal~\cite{vanDriel}. Degenerate two-photon lasing, i.e., emission of two photons into the same mode of the electromagnetic field, has been reported in atomic systems in two exemplary experiments, in the microwave range of the spectrum~\cite{brune87a} and later in the optical range, by pumping the dressed levels of potassium atoms~\cite{Zakrzewski91a,Zakrzewski91b,Zakrzewski91c,gauthier92a}. There is an extensive literature on degenerate two-photon lasing in atoms, some relevant references can be found in~\cite{davidovich87a,zhu87a,zhu88a,boone90a,ninglu90a,lewenstein90a,bay95a,Valcarcel}. Further theoretical works discuss the perspectives of realizing two-photon lasing in semiconductor devices, such as in Refs.~\cite{ironside92a,marti03a,ning04a}. Interest from the solid-state community towards two-photon lasing has been renewed in view of recent experiments, demonstrating two-photon emission in semiconductor systems~\cite{Flissikowski,stufler,hayat08a}, and of the remarkable experimental progress in achieving strong coupling between a single quantum dot and a microcavity mode~\cite{hennessy07a,khitrova06a}. In particular, it is expected that by tailoring the coupling between the quantum dot and the cavity, two-photon emission can be suitably enhanced, while one-photon processes are suppressed~\cite{petrosyan99,vanDriel}.

In this paper we theoretically analyze the possibility of observing two-photon lasing in a semiconductor device consisting of a single quantum dot strongly coupled to the mode of a microcavity.

\begin{figure}[b!]
  \centering
    \includegraphics[width=.8\linewidth]{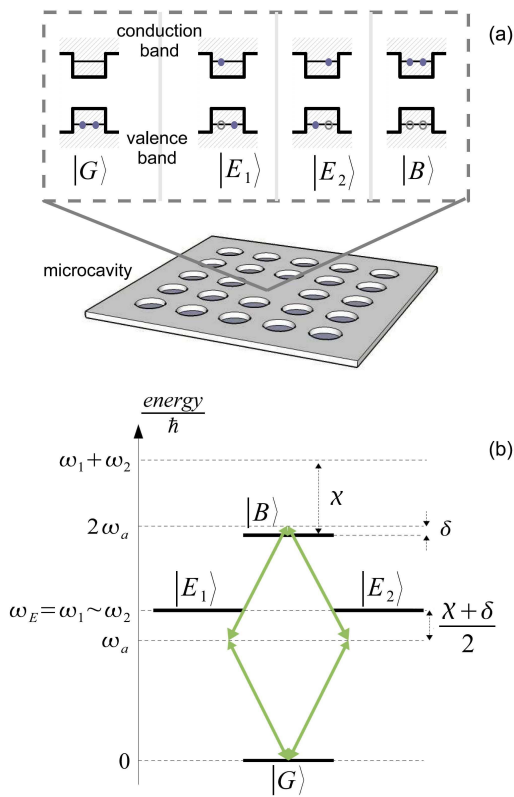}
    \caption{(Color online) (a) A high-$Q$ microcavity coupled to a single quantum dot can be experimentally realized, for instance, by embedding the quantum dot in a photonic crystal~\cite{khitrova06a}. In (b) the ground ($|G\rangle$), excitons ($|E_1\rangle$, $|E_2\rangle$)), and biexciton ($|B\rangle$) energy levels of the quantum dot are displayed, the arrows indicate the coupling due to the resonator mode, the horizontal dashed line shows the frequency of the cavity-mode photon (see Sec.~\ref{Sec:Model} for details on the parameters). The cavity mode is resonantly coupled with the biexciton-ground state transition via two-photon processes. Due to the biexciton-binding energy shift, the exciton states are coupled far-off resonance. In this regime we explore the performance of the system as a two-photon laser.}
    \label{fig:1}
\end{figure}

{The discrete spectrum of a quantum dot (QD) supports a series of optical transitions that can be coupled to one or more cavity modes. Apart from a few localized state in the neutral QD, this nanostructure can also accommodate several carriers in several states, including the continua of states in the wetting layer above the QD electron and hole barriers. When subject to Coulomb interaction, these electronic configurations will form a very rich N-exciton optical spectrum\cite{kaniber,yamaguchi,chauvin,winger09}. In spite of this complicated structure, relevant for many properties, we will focus here on a "minimal" model in which just the lowest energy excitations of the neutral QD are considered as} sketched in Fig.~\ref{fig:1} where the QD is composed by four electronic levels: ground, excitons, and biexciton states (see Fig.~\ref{fig:1}(b)). Previous theoretical works used this level scheme in order to study the emission of polarization-entangled photon pairs from a single quantum dot~\cite{perea05,troiani06b,troiani08,johne08a,pathak09a}. In the present paper, we consider a setup where the same cavity mode couples with the biexciton-exciton and exciton-ground state transitions, setting hence the basic requirements for the emission of two photons into the same mode of the electromagnetic field.

Two-photon gain is reached in the regime where the resonator mode drives resonantly two-photon processes, while one-photon processes are far--off resonance. This is achieved by tuning the cavity frequency to half the biexciton-ground state transition frequency. Due to the binding energy of the biexciton state, this frequency can differ by several meV from the exciton transition, hence in principle offering the ideal conditions for realizing two-photon lasing~\cite{petrosyan99}. In contrast with previous studies, where population inversion is reached by initially preparing the emitter in the excited state of the cascade  emission~\cite{ninglu90a,bay95a,boone90a}, here the quantum dot is continuously pumped, and population inversion may result from the competition between coherent coupling, pumping and losses.

Our study is based on a full quantum mechanical description of the quantum dot and of the cavity field mode. We describe the dissipation caused by the cavity decay, radiative decay and pure dephasing of the quantum dot in the framework of a Born-Markov master equation. In addition, we assume that the quantum dot is continuously pumped by an incoherent field. Incoherent continuous pumping is a typical excitation scheme of semiconductors, where it has proved to be a determining ingredient in the description and achievement of strong coupling~\cite{laussy08a}. The behaviour of the system is studied in the steady state of the dynamics, by analyzing the quantum-dot levels population, the cavity photon statistics, and the spectra of the light intensity at the cavity output as a function of the various parameters. By means of this analysis we identify optimal parameters for observing two-photon gain and we study the transition into the lasing regime. We carry out our analysis with realistic set of parameters, as compared to the semiconductor state of the art, and conclude that semiconductor devices may be suitable for two-photon lasing.

This article is organized as follows. In Sec.~\ref{Sec:Model}, we introduce the theoretical model. The coherent dynamics is { summarized} in Sec.~\ref{sec:coherent}. The incoherent processes are then introduced in Sec.~\ref{sec:incoherent}.  The results are presented in Sec.~\ref{sec:Results}, where we numerically study the steady state of the system, quantum dot and cavity field mode, as a function of the various parameters. Moreover, a semiclassical rate equation for the cavity mode, valid in the limit of a very good cavity and for the initial stages of the dynamics, is obtained in Sec.~\ref{sec:gain}. The conclusions are drawn in Sec.~\ref{sec:Conclusions}, where a detailed discussion on the experimental feasibility of two-photon lasing with a single quantum dot in a cavity is presented. In the appendix the details of the calculations, at the basis of the rate equation in Sec.~\ref{sec:gain}, are reported.

\section{Model}
\label{Sec:Model}

The system we consider is composed by a QD optically coupled to a single mode of the electromagnetic field of a microcavity with a high quality factor $Q$. The relevant states of the quantum dot are the four states of a biexciton system, sketched in Fig.~\ref{fig:1}(b): the ground state $\ke{G}$, at zero energy, the single exciton states $\ke{E_1}$ and $\ke{E_2}$ at frequencies $\omega_1$ and $\omega_2$, respectively, and the biexciton state $\ke{B}$ at frequency
\begin{equation}
  \label{eq:FriMay22193303GMT2009}
  \omega_B=\omega_1+\omega_2-\chi\,,
\end{equation}
where $\chi$ denotes the biexciton energy shift due to the exciton-exciton interaction. The transitions are coupled in a diamond-like configuration of levels with the single mode of the electromagnetic field, which couples to all dipolar transitions of the quantum dot. The master equation for the density matrix $\rho$ of this system, QD and cavity-mode degrees of freedom, reads
\begin{equation}
\label{Master:Eq}
\frac{\partial}{\partial t}\rho=\frac{1}{{\rm i}\hbar}[H,\rho]+{\bf L}\rho\,,
\end{equation}
where $H$ is the Hamiltonian describing the coherent evolution and ${\bf L}$ is the superoperator which describes the effect of noise, dissipation, and incoherent pumping over the coupled system. The Hamiltonian $H$ takes the form
\begin{eqnarray}
  \label{eq:1}
  H&=&\hbar\omega_a a\da a+\sum_{j=1,2}\hbar \omega_j \kb{E_j}{E_j}+\hbar \omega_B\kb{B}{B}\nonumber\\
  &&+\hbar g\sum_{j=1,2} \pq{\pt{\kb{G}{E_j}+\kb{E_j}{B}}a\da  + {\rm H.c.}}\,,
\end{eqnarray}
with $\omega_a$ the cavity frequency, $a$ and $a^\dagger$ the annihilation and creation operators of a cavity photon,
{
and g is the vacuum Rabi frequency, which is here assumed to be equal for all transitions. It is important to note that we make this approximation in order to simplify the discussions in the following Sections. Typically, the couplings between each of the two excitons and the cavity mode differ, as they depend on the QD and cavity geometry (see, for instance, Ref. \onlinecite{winger08} and references therein). However, considering different coupling constants for the four transitions in the system is not significant for the two-photon lasing mechanism, as we show in Appendix B. There, we carried out a systematic analysis of this issue and we found that, within some reasonable restrictions for the parameters, the results are qualitatively the same as the ones we present in the main text using homogeneous couplings for all the transitions. We also note that this simplifying assumption is relevant for the coherent dynamics, as it introduces additional symmetries in the level couplings, see for instance Refs.~\onlinecite{Morigi02,machnikowski08a}.

As we are interested in two-photon processes, we will focus on configurations where the intermediate, exciton levels are far-off resonantly coupled to the radiation, while ground state and biexciton state are resonantly coupled by two-photon processes. In what follows we neglect the energy difference between the exciton states, assuming that the frequency difference $\abs{\omega_1-\omega_2}$  is much smaller than all the other characteristic frequencies of
the system. In this limit, which is representative enough of more general situations, see App. B, we define the exciton frequency $\omega_E\equiv\omega_1\simeq\omega_2$.}
In the frame rotating at the cavity frequency, the Hamiltonian becomes
\begin{eqnarray}
  \label{eq:1:0}
  H^{\prime}&=&\hbar\Delta\sum_{j=1,2}\kb{E_j}{E_j}+\hbar \delta\kb{B}{B}\nonumber\\
  &&+\hbar g\sum_{j=1,2} \pq{\pt{\kb{G}{E_j}+\kb{E_j}{B}}a\da  + {\rm H.c.}}\,,
\end{eqnarray}
where we have introduced the one-photon detuning
\begin{equation}
\Delta=\frac{\chi+\delta}{2}\,,
\end{equation}
and where $\delta$ is the biexciton detuning, defined as
\begin{equation}
  \label{det:2P}
  \delta=\omega_B-2\omega_a\,.
\end{equation}

We remark that our model scheme assumes that the same cavity mode couples to both biexciton-exciton state transitions, $|B\rangle\leftrightarrow|E_j\rangle$, and to both exciton-ground state transitions, $|E_j\rangle\leftrightarrow|G\rangle$. This choice is made in order to realize a degenerate two-photon laser, and requires a cavity where the two polarization modes have well separated frequencies, as well as a proper definition of the quantization axis (which is naturally set by the geometric asymmetry of the quantum
dot).

\subsection{Coherent two-photon processes}
\label{sec:coherent}

We first focus on the coherent dynamics of the QD coupled with the microcavity field, discarding for the moment any source of noise, and identify the region of parameters in which two-photon absorption and emission are significantly enhanced over single-photon processes.

\begin{figure}[htb]
  \begin{center}
      \includegraphics[width=\linewidth]{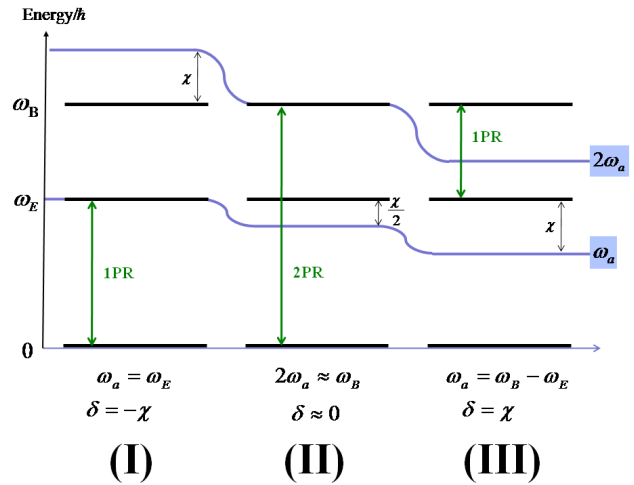}
        \caption{(Color online) Energy levels of the QD for different cavity-mode frequencies, when the biexciton is bound ($\chi>0$). In (I) the cavity mode is resonant with the transition $\ket{G}\leftrightarrow\ke{E}$, namely, $\delta=-\chi$ ($\omega_E=\omega_a$). In (II) the $\ke{G}\leftrightarrow\ke{B}$ transition couples resonantly with two photons of the cavity mode, $\delta\simeq0$ ($2\omega_a\simeq\omega_B$). In (III) the cavity mode is resonant with the exciton-biexciton transition $\ke{E}\leftrightarrow\ke{B}$ ($\delta=\chi$, corresponding to $\omega_a=\omega_B-\omega_E$).}
\label{fig:levels-change}
  \end{center}
\end{figure}

We denote here by \emph{two-photon oscillations} the coherent exchange of photons in pairs between the cavity mode and the quantum dot. This is found when excitations and de-excitations of the transition $|G\rangle\leftrightarrow |B\rangle$ correspond to resonant absorption and emission, respectively, of two photons of the cavity mode (corresponding to setting $2\omega_a\simeq\omega_B $, i.e. $\delta\simeq 0$), while the single-exciton states remain effectively unoccupied. The latter condition is achieved when the single-photon detuning is much larger than the coupling to the cavity mode, $\chi\gg g\sqrt{\langle \hat n\rangle}$, where $\langle \hat n\rangle$ is the mean intracavity photon number. In this limit, we expect Rabi oscillations between the ground and the biexciton state with maximum amplitude, such that periodically one finds population inversion in the transition. The possible { relevant cases} are summarized in  Fig.~\ref{fig:levels-change}, { where  the cavity-mode frequency is shown with respect to the QD levels} when the biexciton is bound. In particular, the two-photon resonance regime is sketched in Fig.~\ref{fig:levels-change}(II) (see the caption for details).

Let us now describe in more detail the dynamics of this system at two-photon resonance. In order to proceed with our analysis we introduce the basis of states $|\xi,n\rangle$, where $\xi=G,E_1,E_2,B$ denote the QD states and $n$ is the number of cavity photons. Hamiltonian~(\ref{eq:1}) couples the closed set of states $|G,n\rangle$, $|E_j,n-1\rangle$, $|B,n-2\rangle$. For $|\delta|\ll g\sqrt{\langle \hat n\rangle}\ll\chi $ one can eliminate the exciton states from the coherent dynamics of the ground and biexciton states. By applying the Schrieffer-Wolf transformation~\cite{schrieffer66a} one obtains an effective hamiltonian for the reduced Hilbert space of the QD states $\{|G\rangle,|B\rangle\}$ and for the cavity mode states, which reads
\begin{eqnarray}
  \label{Heff}
H_{\rm eff}&=&\hbar(\delta+\delta_{2P})\kb{B}{B}
\\&&
  +\hbar g_{\rm eff}\pt{a^2\kb{B}{G}+{a\da}^2\kb{G}{B}}\,,\nonumber
  \end{eqnarray}
where \begin{equation}
\label{geff} g_{\rm eff}=-\frac{4g^2}{\chi}
\end{equation}
scales the amplitude of the two-photon Rabi oscillations between states $|G\rangle$ and $|B\rangle$, and
\begin{equation}
\delta_{2P}=\frac{4g^2}{\chi}
\end{equation}
is the frequency shift on the two-photon transition, resulting from the coupling to the intermediate, exciton states. Correspondingly the ground-biexciton transition $|G\rangle\leftrightarrow |B\rangle$ and cavity field are in two-photon resonance when
\begin{equation}\label{delta}
\delta=-\delta_{2P}.
\end{equation}
This frequency shift is independent of the number of photons and has been discussed in the literature on two-photon lasers as self-induced frequency pulling term~\cite{davidovich87a}. Hamiltonian~(\ref{Heff}) has been obtained assuming that $\chi$ is the largest parameter, so that the intermediate states are coupled far-off resonance from ground and biexciton state by the cavity-mode photons~\footnote{It must be kept in mind that the coupling with the intermediate, single exciton states can be discarded provided that the frequency shifts  $|\delta_{2P}|\ll g\sqrt{\av{\hat n}}$.}. We note that the amplitude of the two-photon coupling, Eq.~(\ref{geff}), is inversely proportional to $\chi$. Hence, by choosing large single--photon detunings, of the order of  $\chi$, the effect of the intermediate states is negligible. This choice implies, however, that  the period of the two-photon Rabi oscillation, proportional to $\chi$, becomes very large, so that { the effect of noise and dissipation}, neglected in this analysis, may become relevant.

\begin{figure}[htb]
  \begin{center}
   \includegraphics[width=\linewidth]{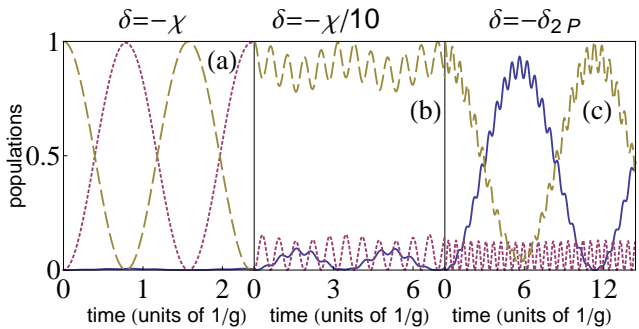}
       \caption{(Color online) Populations of the QD levels $p_{\xi}(t)$ as a function of time, in units of $1/g$, for the initial state $\ket{G,2}$, obtained by evaluating numerically the equation $p_{\xi}(t)=\sum_n\abs{\br{\xi,n}{\rm e}^{-{\rm i}H t/\hbar}\ke{G,2}}^2$. The curves correspond to $p_{B}$ (solid-blue line), $p_{E}=p_{E_1}+p_{E_2}$ (dotted-purple) and $p_{G}$ (dashed-brown). In (a) $\delta=-20g$, (b) $\delta=-2g$, (c) $\delta=-\delta_{2P}$, while $\chi=20g$ in all three cases. See also Fig.~\protect\ref{fig:levels-change}.}
    \label{fig:3}
  \end{center}
\end{figure}

In Fig.~\ref{fig:3} the time-dependent behavior of the population of the QD levels is evaluated numerically for different values of the cavity frequencies and assuming that at $t=0$ the system is in level $|G,2\rangle$. The curves are found by integrating numerically Eq.~(\ref{Master:Eq}), setting the incoherent term to zero. The biexciton binding energy is fixed at $\chi=20 g$. The Figure displays the time evolution of the population of the ground, intermediate, and biexciton levels for different values of the detuning~$\delta$ (and therefore the cavity frequency $\omega_a$), corresponding to the situations sketched in Fig.~\ref{fig:levels-change}. In Fig.~\ref{fig:3}(a) the cavity is resonant with the electronic transition between ground and single-exciton states, and we observe single-photon Rabi oscillations. In (b) the value of the cavity frequency $\omega_a$ is far-off resonance from any electronic transition, and one observes that the population of the initial state is almost unit at all times. When the two-photon resonance condition~(\ref{delta}) is fulfilled, the coherent two-photon exchange $|G,n\rangle\leftrightarrow |B,n-2\rangle$ is realized with maximum amplitude, as we can see in Fig.~\ref{fig:3}(c). Transfer of population between ground and biexciton states occurs with a period of time of the order of $t_{\rm 2P}\simeq \pi/(2g_{\rm eff})$ (for $n=2$), that corresponds to the time estimated from the dynamics predicted by Hamiltonian~(\ref{Heff}), and found in perturbation theory by eliminating the intermediate levels. Note that the residual population of the single exciton state, which is observed in this plot, can be further suppressed by increasing the biexciton binding energy $\chi$, at the expenses of increasing $t_{\rm 2P}$. The curve, corresponding to the ground state population,  gives also the probability of finding two photons inside the resonator.

\begin{figure}[htb]
  \begin{center}
    \includegraphics[width=\linewidth]{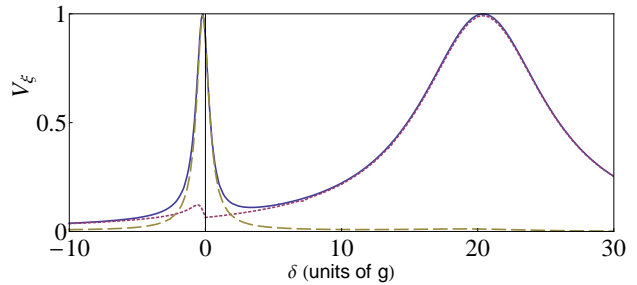}
    \caption{(Color online) Amplitude of oscillations of the population of the QD states, $V_\xi={\rm max}\pg{p_{\xi}(t)}-{\rm min}\pg{p_{\xi}(t)}$ as a function of $\delta$, in units of $g$, with $\chi=20g$ and initial state $|B,0\rangle$. The curves correspond to $V_B$ (solid-blue line), $V_E$ (dotted-purple) and $V_G$ (dashed-brown).}
    \label{fig:4}
  \end{center}
\end{figure}

Figure~\ref{fig:4} displays the maximum value of the amplitude of the oscillation of the populations of the QD states as a function of $\delta$ for $\chi=20 g$, assuming that the initial state is $|B,0\rangle$. Unit amplitude is found when the cavity is either resonant with the exciton-biexciton transition ($\delta=\chi$, see Fig.~\ref{fig:levels-change} (III)) or with the ground-biexciton transition ($\delta\simeq 0$, see Fig.~\ref{fig:levels-change} (II)). The widths of the two peaks are related to the strength of the excitons photon coupling. In particular, the two-photon peak is located at the two-photon resonance, Eq.~(\ref{delta}), and is much narrower than the single-photon resonance because of the second-order coupling between ground and biexciton state.

\subsection{Incoherent processes}
\label{sec:incoherent}

We now include the incoherent processes in the dynamics of QD and cavity mode. In the present model, they are due to the coupling of system with the environment constituted by the external modes of the electromagnetic field and the phononic modes of the sample in which the QD is embedded, and are described by the Liouvillian ${\bf L}$ in the master equation, Eq.~(\ref{eq:1}). This includes the decay of the cavity-mode photons into the external modes of the electromagnetic field at rate $\kappa$, decay of the exciton and biexciton states at rate $\gamma$ and $\gamma_B$, respectively (which include radiative and non-radiative losses, the latter due to electron-phonon interactions), and pure exciton dephasing at rate $\Gamma_d$, due to interaction with the bulk phonons, which we assume here to be Markovian~\cite{Imamoglu2004}. In addition, the QD is pumped continuously. Typically, electron-hole pairs are created in high energy states by electrical injection or optical pumping. Then they relax to the excitonic levels of the QD producing an effective incoherent source of particles with injection rate which we denote here by~$P$. The Lindblad term ${\bf L}$ in Eq.~(\ref{eq:1}), describing these processes, takes the form
\begin{eqnarray}\label{Lincoh}
{\bf L}\rho&=&\frac{\kappa}{2}{\cal L}(a)\rho  \nonumber\\
    &&+\sum_{j} \pq{\frac{\gamma}{2}{\cal L}\pt{\kb{G}{E_j}}+\frac{\gamma_B}{4}{\cal
        L}\pt{\kb{E_j}{B}}}\rho
    \nonumber\\
    &&+\frac{\Gamma_d}{2}\sum_j{\cal L}\pt{\kb{E_j}{E_j}}\rho +\Gamma_d{\cal L}\pt{\kb{B}{B}}\rho\nonumber\\
    && +\frac{P}{2}\sum_{j} \pq{{\cal L}\pt{\kb{E_j}{G}}+{\cal
        L}\pt{\kb{B}{E_j}}}\rho\,,
\end{eqnarray}
where we have used the definition
\begin{eqnarray}
    {\cal L}(x)\rho=2 x\rho x\da-x\da x\rho-\rho x\da x\,,
\end{eqnarray}
with $x$ an operator acting on the Hilbert space of cavity mode and QD. In the following we assume for simplicity that
\begin{eqnarray}
    \gamma_B=2\gamma\,.
\end{eqnarray}
We finally note that in the literature pure dephasing has been sometimes described with non-Markovian type of Liouvillians~\cite{Calarco,Kuhn,machnikowski08a}. It is indeed expected that non-Markovian corrections will give rise to non-linearities in the phonons-cavity coupling. In this paper we describe dephasing as Markovian process~\cite{Imamoglu2004}, as our scope is to identify the parameter regime where two-photon lasing can be observed.

The incoherent dynamics in general competes with the coupling with the cavity mode. Noise and dissipation, hence, will tend to suppress two-photon coherent exchange between QD and cavity mode. Nevertheless, quantum effects, at the levels of few photons, will affect relevantly the dynamics provided that the coupling strength of the QD with the cavity mode exceeds the loss rates. An important parameter, determining whether the system is in the strong coupling regime, is the cooperativity. It is defined as~\cite{carmichael,kimble98}:
\begin{equation}
C=\frac{2 g^2}{\kappa\gamma'}\,,
\end{equation}
where $\gamma'=\gamma+\Gamma_d$. In absence of incoherent pumping strong coupling is warranted when $C>1$.

\section{Two-photon lasing}
\label{sec:Results}

In this section we study the steady-state properties of the system evaluated by solving numerically the master Eq.~(\ref{Master:Eq}) with the Lindblad term (\ref{Lincoh}) as a function of the physical parameters. We first focus on the case in which dephasing is set to zero, in order to single out the role of the various incoherent processes on the lasing dynamics, and then consider the effect of dephasing within the description in Eq.~(\ref{Lincoh}).
We finally discuss a specific parameter regime where an analytic solution can be found for the dynamics in a very good cavity, and allows us to identify { a condition when} two-photon gain exceeds one-photon gain in the initial stages of the dynamics. Comparison with the numerical results shows that this condition does not necessarily imply that the system accesses the two-photon lasing regime.

We remark that previous theoretical works studying two-photon lasing in cavities focussed on setups where atoms in the upper excited state were pumped into the cavity. In that case, population inversion on the two-photon transition is warranted by the state preparation of the atoms entering the cavity, and the rate at which they cross the resonator is one important parameter controlling whether two-photon lasing is found~\cite{ninglu90a,bay95a,boone90a}. In this work we consider a single dot inside a resonator whose levels are continuously pumped by external, incoherent fields. Population inversion, if found, hence results from the competition between the pump, the loss and the emission processes.

\subsection{Steady state and photon statistics}
\label{steadystate}

\begin{figure}[!ht]
\begin{center}
   \includegraphics[width=8.5cm]{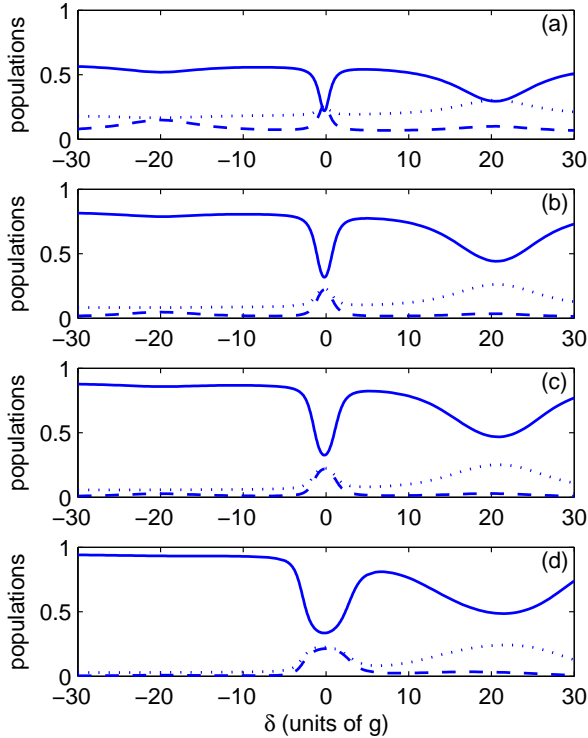}
   \caption{(Color online) Steady state populations of the QD energy levels as a function of the cavity detuning $\delta$ (in units of $g$) for  $\kappa=0.1g$, $\Gamma_d=0$, $\gamma=0.05g$, $\chi=20g$ and pump rate (a) $P=0.16g$, (b) $P=0.5g$, (c) $P=0.8g$, (d) $P=1.8g$. The curves correspond to the occupation of the states $|B\rangle$ (solid line), $|E_j\rangle$ (dotted line), and $|G\rangle$ (dashed line). Two-photon resonance is at $\delta=-\delta_{2P}\sim 0$. At $\delta=20g$ transition $|B\rangle\leftrightarrow|E_j\rangle$ couples resonantly with the cavity mode.}
  \label{fig:ssQD}
  \end{center}
\end{figure}

\begin{figure}[!ht]
\begin{center}
   \includegraphics[width=8.5cm]{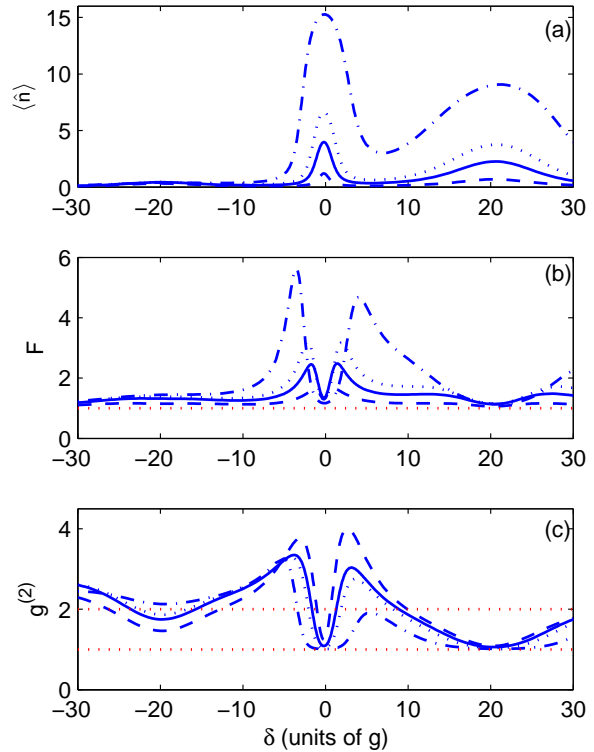}
   \caption{(Color online) (a) Average photon number $\av{\hat n}$, (b) Fano factor $F$, and (c) $g\al{2}(0)$  as a function of the cavity detuning $\delta$ (in units of $g$) for $\kappa=0.1g$, $\Gamma_d=0$, $\gamma=0.05g$, $\chi=20g$ and  pump strengths $P=0.16g$ (dashed lines), $P=0.5g$ (solid lines), $P=0.8g$ (dotted lines), $P=1.8g$ (dotted-dashed lines). The horizontal dotted lines indicate (b) the value $F=1$, (c) the values $g^{(2)}(0)=1$ (Poisson distribution) and $g^{(2)}(0)=2$ (thermal distribution). }
  \label{fig:ss}
  \end{center}
\end{figure}

We first study the steady-state population of the QD-electronic states as a function of the cavity frequency and of the pump strength. As is visible in Fig.~\ref{fig:levels-change}, the cavity-QD system possesses three resonances: the two-photon resonance at $\delta=-\delta_{2P}$, when the transition $|B\rangle\leftrightarrow |G\rangle$ is resonantly driven, and the one-photon resonances at $\delta=\chi$ and $\delta=-\chi$, when the transitions $|B\rangle\leftrightarrow |E_j\rangle$ and $|E_j\rangle\leftrightarrow |G\rangle$, respectively, are resonantly driven.

When the cavity is far--off resonance from these transitions, then the steady state of the electronic levels is determined by the incoherent processes of exciton pumping and decay described in Eq.~(\ref{Lincoh}). Denoting by $p_j=\langle j|\rho|j\rangle$ the population of the electronic state $|j\rangle$ ($j=G,B,E_1,E_2$), in this limit they take the form $p_G=\gamma^2/(P+\gamma)^2$, $p_B=P^2/(P+\gamma)^2$, and $p_{E_1}=p_{E_2}=\gamma P/(P+\gamma)^2$, showing that the biexciton state is occupied with almost unit probability when $P$ exceeds the radiative decay rate. At the resonances, this behaviour is significantly modified by the competition between the incoherent processes and the strong coherent coupling to the cavity mode.

Figure~\ref{fig:ssQD} displays the value of the populations at steady state as a function of the two-photon detuning $\delta$, where here $\delta=\pm 20g$ corresponds to the one-photon resonances. The subplots from top to bottom are calculated for increasing values of the pumping rate $P$. Here, one observes depletion of the biexciton population at $\delta\sim 0,\chi$, which is more enhanced at two-photon resonance. At $\delta\sim \chi$, when the cavity mode drives resonantly the biexciton-exciton transition, depletion of the biexciton state population is accompanied by a significant increase of the intermediate levels occupations, while the ground state remains essentially empty. Photon emission on the biexciton-exciton transition is here expected. At two-photon resonance one observes a significant population of the ground state, while the (small) increase of the intermediate states populations further indicates that the dominating processes are emission of photons in pairs from the biexciton to the ground state level. Increase of the pump strength leads to stationary population inversion between the biexciton and the other levels, and to a broadening of the interval of values of $\delta$ in which population depletion is observed.

\begin{figure}[!t]
\begin{center}
   \includegraphics[width=8.5cm]{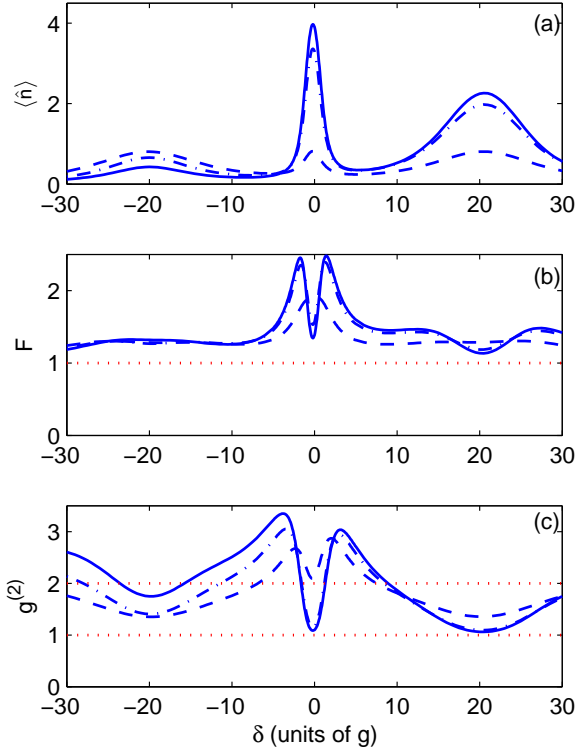}
   \caption{(Color online) Same as in Fig.~\ref{fig:ss}, but for a fixed value of the pump rate, $P=0.5g$, and for different values of the radiative decay rate: $\gamma=0.05g$ (solid line), $\gamma=0.1g$ (dot-dashed line), $\gamma=0.5g$ (dashed line).}
  \label{fig:ss2}
  \end{center}
\end{figure}

We now analyze the cavity field properties as a function of $\delta$. Figure~\ref{fig:ss} displays (a) the mean number of photons, (b) the Fano factor, which is the ratio of the  variance of what is measured to that for a Poisson distribution with the same mean photon number, and it is defined as~\cite{mandel95a}
\begin{eqnarray}
    F=\frac{\av{{\hat n}^2}-\av{\hat n}^2}{\av{\hat n}}\,,
\end{eqnarray}
with $\hat n=a\da a$ the photon number operator, and (c) the intensity-intensity correlation function at zero delay,
\begin{eqnarray}
    g^{(2)}(0)=\frac{\av{{\hat n}^2}-\av{\hat n}}{\av{\hat n}^2}\,,
\end{eqnarray}
quantifying the probability of two-photon coincidence detection. The curves are calculated at relatively small values of the radiative decay and zero dephasing rate, and for different values of the pump strength. The photon number exhibits two peaks, at two-photon and one-photon resonance, whereby the maximum at the two-photon resonance is higher than at one-photon resonance. This behavior is enhanced as the pump strength is increased. Moreover, while in general the peak at two-photon resonance is narrower than the peak at one-photon resonance, it broadens significantly for large pump powers. The photon-number fluctuations, on the other hand, decrease at two- and one-photon resonance. Fano factor and $g^{(2)}(0)$ approach unity at $\delta=\chi$ for large pump rates, while they remain larger than unity at $\delta=0$. Close to the minimum at two-photon resonance large fluctuations in the photon number are observed. Finally, far from the one- and two-photon resonances the second-order correlation functions approaches the value $g^{(2)}(0)\sim 2$, which indicates a thermal character of the field.

\begin{figure}[!t]
\begin{center}
   \includegraphics[width=8.5cm]{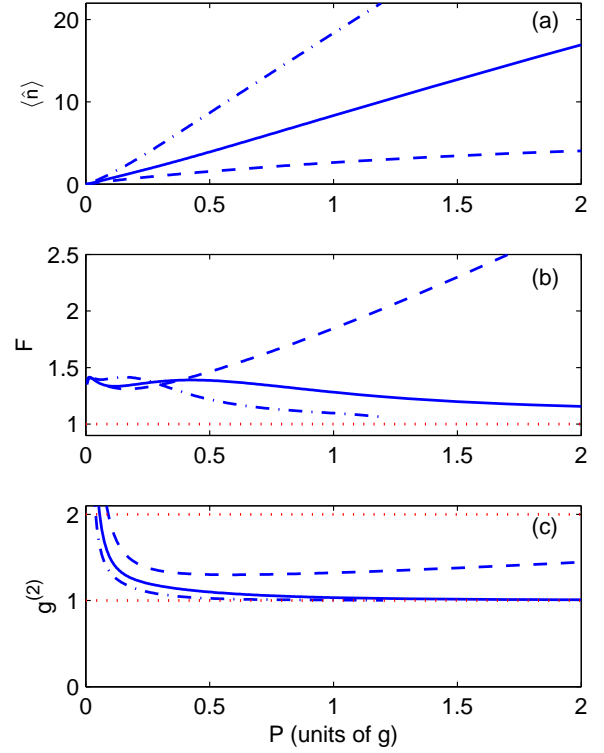}
      \caption{(Color online)(a) Average photon number, (b) Fano factor, and (c) $g^2(0)$ as a function of the pumping rate $P$ (in units of $g$) for $\delta=0$, $\Gamma_d=0$, $\gamma=0.05g$, $\chi=20g$, and different values of the cavity decay rate: $\kappa=0.2g$ (dashed line), $\kappa=0.1g$ (solid line), and $\kappa=0.05g$ (dot-dashed line).}
  \label{fig:ssF}
  \end{center}
\end{figure}

\begin{figure}[!t] \begin{center}
\includegraphics[width=8cm]{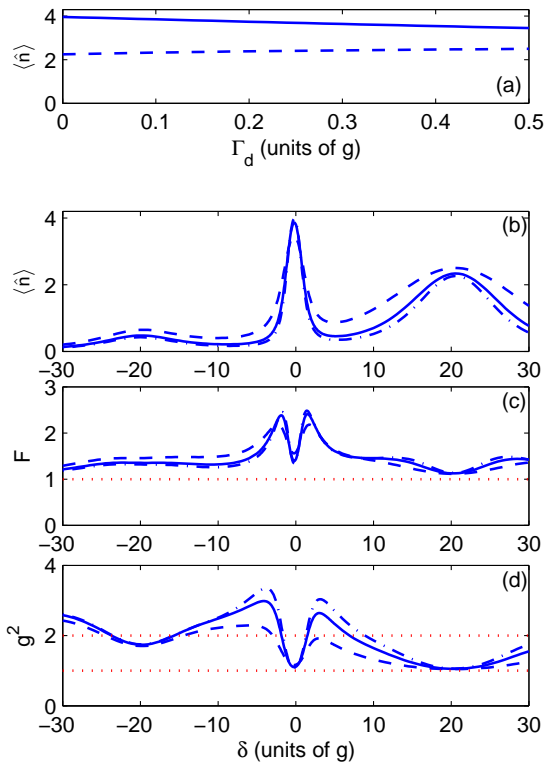}
\caption{(Color online) (a) Cavity-field average photon number $\av{\hat n}$ at steady state as a function of the dephasing rate $\Gamma_d$ in units of $g$. The solid line corresponds to $\delta=0$ (two-photon resonance), and the dashed line to $\delta=\chi$ (one-photon resonance). (b) Average photon number, (c) Fano factor, and (d) $g^{(2)}(0)$ as a function of the two-photon detuning $\delta$ in units of $g$. The dot-dashed line corresponds to $\Gamma_d=0$, the solid line to $\Gamma_d=0.1g$, the dashed line to $\Gamma_d=0.5g$.
The other parameters are $P=0.5g$, $\gamma=0.05g$ and $\kappa=0.1g$. }\label{fig4}
\end{center}
\end{figure}

The effect of radiative decay on the properties of the cavity field is analyzed in Fig.~\ref{fig:ss2}. Here, we observe that when the decay rate is increased, the number of photons at one- and two-photon resonance decreases (while the photon number at the one-photon resonance of the exciton-ground state transition increases), and correspondingly the photon-number fluctuations increase. When $P\sim \gamma$, in particular, the Fano factor is maximum at the two-photon resonance, see Fig.~\ref{fig:ss2}(b). Correspondingly, $g\al{2}(0)\sim 2$, indicating that the cavity field approaches a thermal distribution.

A further important parameter is the rate at which the cavity field leaks out of the resonator. In Fig.~\ref{fig:ssF} we study the behavior of the steady state photon statistics as a function of the exciton pumping rate for various values of the field decay rate. While it is obvious that, increasing the pump and decreasing the cavity loss rate, the number of photons inside the cavity increases, the behaviour of the photon-number fluctuations is interesting. We observe that $g^{(0)}>1$ ($F>1$) for all values of $\kappa$ and of $P$ here considered, indicating that the field is superpoissonian. Moreover, the Fano factor increases with increasing pump rates at $\kappa=0.2g$, while it decreases to values which approach unity for $\kappa=0.05g$ and $0.1g$. It is interesting to note that for a certain interval of values of $P$, which are much smaller than the cavity coupling strength $g$, the Fano factor evaluated at $\kappa=0.05g$ and $\kappa=0.1g$ exhibits a broad maximum. We note that values of the Fano factor $F>1$ are expected in two-photon lasers, see for instance Ref.~\cite{ninglu90a,bay95a}. A direct comparison with the values found in the literature and the ones we obtain is not trivial, as differing from the systems in~\cite{ninglu90a,bay95a} the setup we consider has relatively large loss rates $\kappa$ and a different pumping mechanism.

The effect of dephasing on the state of the cavity field is studied in Fig.~\ref{fig4}. Figure~\ref{fig4}(a) displays the mean intracavity photon number as a function of the dephasing rate, showing that $\langle \hat n\rangle$ decreases at two-photon resonance as $\Gamma_d$ is increased, while it increases with $\Gamma_d$ at one-photon resonance. Figures~\ref{fig4}(b)-(d) analyzes this behaviour in the photon statistics, showing that the one-photon resonance as a function of $\delta$ is higher and broader for the largest value of the dephasing rate here considered, $\Gamma_d=0.5g$. We note, however, that this effect is observed only when the dephasing rate exceeds the radiative decay rate, as we assume in this specific case.

\subsection{Spectrum of the cavity field}
\label{sec:Spectra}

\begin{figure*}[!t]
\begin{center}
   \includegraphics[width=17cm]{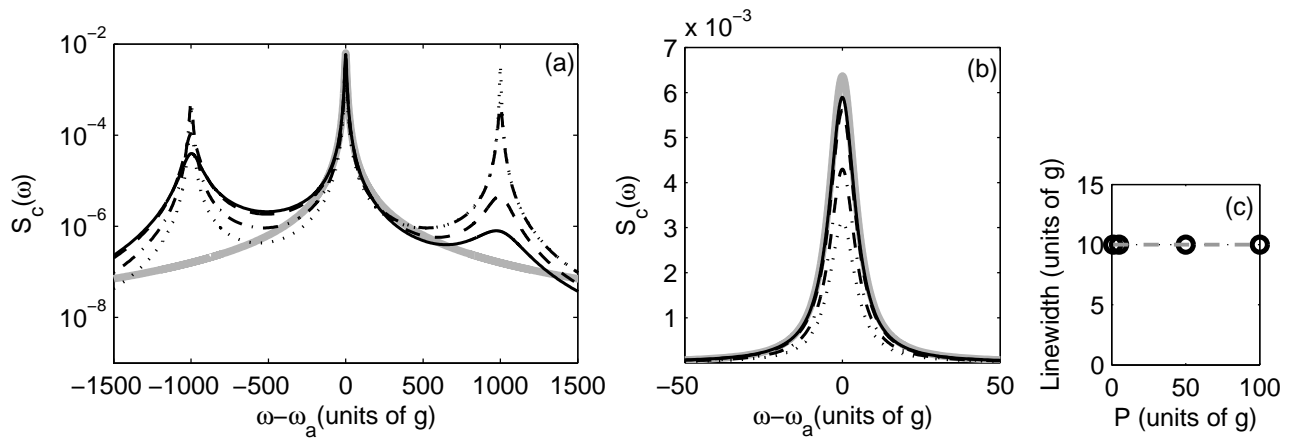}
         \caption{\label{spectrum} (a) Spectrum of the cavity field intensity at the output for different values of the pump rate $P$ with $\kappa=10g$, $\delta=0$, $\chi=2000g$, $\gamma=5g$. The curves refer to $P=g$ (dotted line), $P=5g$ (dotted-dashed line), $P=50g$ (dashed line), $P=100g$ (solid line). The plot in (b) shows the curves in the interval of frequency close to $\omega-\omega_a=0$ and plotted in linear scales. The thick gray curves in (a) and (b) indicate the (Lorentzian) emission line of the cavity in absence of the coupling with the quantum dot. The plot (c) shows the linewidth at half height of the peaks in (b) as a function of P. The horizontal dashed line indicates the value of $\kappa$.}
    \end{center}
  \end{figure*}

\begin{figure*}[!t]
\begin{center}
       \includegraphics[width=17cm]{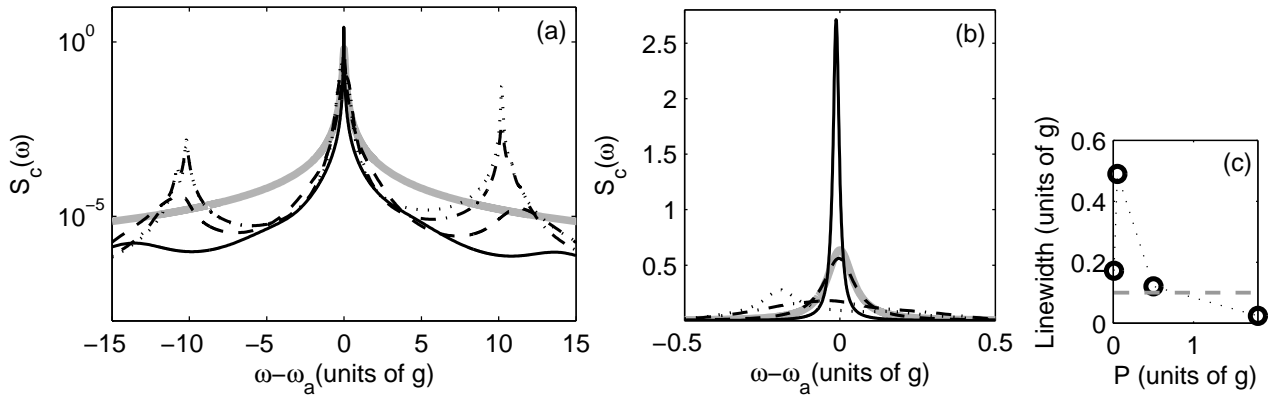}
      \caption{\label{spectrum2} Same as Fig.~\ref{spectrum}, but with $\kappa=0.1g$, $\delta=0$, $\chi=20g$, $\gamma=0.05g$ and $P=0.01g$ (dotted line), $P=0.05g$ (dotted-dashed line), $P=0.5g$ (dashed line), $P=1.8g$ (solid line).}    \end{center}
\end{figure*}

We now focus on the spectrum of the field at the cavity output, which is evaluated from the formula
\begin{equation}
  \label{Spectrum}
  S_c(\omega)=\frac{\kappa}{\pi
    \mean{\ud{a}a}_{\infty}}\Re\int_{0}^{\infty}\langle\ud{a}(\tau)a(0)\rangle_{\infty}
  e^{i\omega\tau}d\tau\,,
\end{equation}
using the quantum regression theorem~\cite{walls94a,bienert07a,delvalle08a}, and where we assumed that the input field is in the vacuum, consistently with master equation~(\ref{Master:Eq}). We define $\langle\cdot\rangle_{\infty}\equiv {\rm Tr}\{\cdot \rho_{\infty}\}$, where $\rho_{\infty}$ is the density matrix of QD and cavity field at steady state. Below we analyze the spectrum when the system is driven at $\delta=0$.

Figure \ref{spectrum} displays the spectrum at large loss rates and small cooperativities. We note that it exhibits three main peaks at the bare system frequencies: one at the cavity frequency $\omega=\omega_a$, a second at the ground-exciton transition frequency and a third one at the frequency of the exciton-biexciton transition.
In particular, the central curves are Lorentzian with linewidth $\kappa$, as for the cavity field in absence of the QD, see Fig.~\ref{spectrum} (c).
In this regime, hence, the spectrum provides a spectroscopic measure of the characteristic frequencies of the system and of their corresponding linewidths. We note, however, that due to the small coupling the exciton resonances are two orders of magnitude smaller than the central peak. The effect of increasing the pump power, nevertheless, is to increase the number of cavity photons at the frequency of the biexciton-exciton transition, while the number of photons at the frequency of the exciton-ground state transition becomes smaller.

Significant differences from this behavior are observed at large cooperativity, with coupling strengths overcoming the loss rates. Figure~\ref{spectrum2} displays the spectrum evaluated as a function of the pump power and large cooperativities. The central peak is dominant and, by increasing the pump, moves from being centered at $\omega=\omega_a-\delta_{2P}$ towards $\omega=\omega_a$. As the pump power is increased, moreover, the exciton resonances both decrease, while the linewidth of emission line at the central peak becomes narrower, as it is visible in Fig.~\ref{spectrum2} (c), thereby exhibiting a typical signature of lasing at this frequency.

\subsection{Non-classical properties of the cavity field}\label{squeezing}

The analysis done so far shows several features that can be interpreted altogether so that the system accesses two-photon lasing at large cooperativity and sufficiently large pump power. In fact, at two-photon resonance and sufficiently large pumping rate we observe population inversion on the biexciton-ground state transition (as well as on the biexciton-exciton transition). Moreover, at these values the statistics of the light is superpoissonian, exhibiting values of the Fano factor which are consistent with the ones expected for a two-photon laser~\cite{ninglu90a,bay95a}. In addition, for these parameters we observe that the cavity spectrum has intense emission and line narrowing at the cavity frequency.

Two-photon lasers are expected to be nonclassical light sources, however, the analysis presented so far does not show clear nonclassical features of the cavity field. In the literature, nonclassical features of degenerate two-photon lasers are usually associated with squeezing of a quadrature at the cavity output. When the system is pumped by atoms prepared in the excited state, squeezing of the cavity field can be observed if a seed is put in the system. If, for instance, the cavity field has been initially prepared in a coherent state, squeezed light is observed when incoherently pumping the system for a transient time, of the order of typical time scale determining phase diffusion~\cite{davidovich87a,bay95a}. In this work we check that such behavior is observed also in the setup we consider, namely, a single QD continuously pumped and strongly coupled to the cavity mode. The results are obtained by numerical integration of the initial master equation~(\ref{Lincoh}).

\begin{figure}[!th] \begin{center}
      \includegraphics[width=8cm]{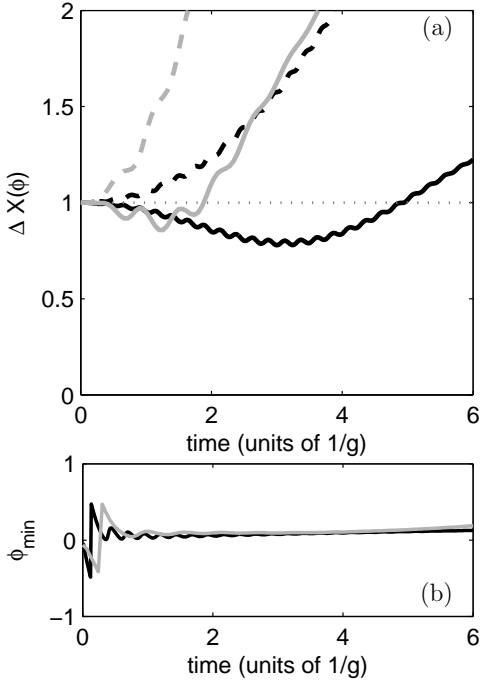}
      \caption{(a) Time evolution of the variance ($\Delta X(\phi)=\av{X^2}-\av{X}^2$) of two orthogonal quadrature of the field, defined as $X=a {\rm e}^{-{\rm i}\phi}+a\da {\rm e}^{{\rm i}\phi}$ (the horizontal dotted line indicates the shot-noise limit). The solid (dashed) lines corresponds to the quadrature $\Delta X(\phi)|_{\rm min}$ ($\Delta X(\phi+\pi/2)|_{\rm max}$) with minimum (maximum) value at the given time. The corresponding phase $\phi$ of the quadrature with minimum variance is plotted in (b). The field is initially prepared in a coherent state with average photon number $\av{\hat n}=4$. The black lines are found for $P=0.01g$, $\gamma=0.01g$ and $\kappa=0.01$, whereas the gray lines are found for larger rates of incoherent processes: $P=0.1g$, $\gamma=0.05g$ and $\kappa=0.02$. The other parameters are $\chi=50g$, $\delta=-\delta_{2P}$.  }
\label{fig5}
\end{center}
\end{figure}

Figure~\ref{fig5} displays the time evolution of the variance of the field quadrature which possesses minimum variance at all times. For sufficiently short times, the field becomes squeezed.  At longer time the squeezing disappears because of the phase diffusion induced by the incoherent processes. The interval of time in which the squeezing is visible is shorter for larger rates of the incoherent processes.

\subsection{A rate equation for the cavity field dynamics}
\label{sec:gain}

The possibility of realizing media exhibiting two-photon gain have been extensively discussed in the literature. Theoretically, one typically estimates gain associated with two-photon processes using a semiclassical analysis. Two-photon lasing is expected when the two-photon gain exceeds the one-photon gain~\cite{ning04a}.

\begin{figure}[!th] \begin{center}
    \includegraphics[width=6cm]{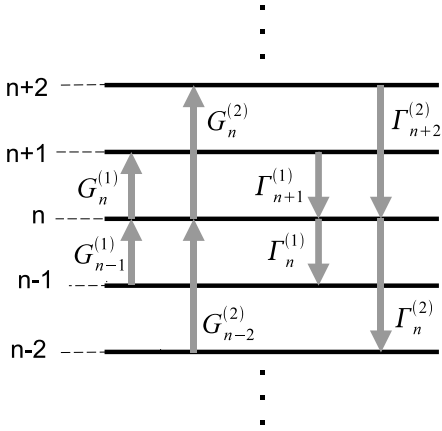}
    \caption{Sketch of the processes considered in the rate
      equation~(\ref{rateEq1}). The horizontal lines indicate the energy levels of the cavity mode, the vertical arrows indicate the flow of
      probability in and out of the energy level at $n$ photons with the corresponding rate.  }
    \label{figRate1}
  \end{center}
\end{figure}

In this section we derive a rate equation for the cavity-photon state occupations. The rate equation is obtained assuming that the intermediate (exciton) levels are coupled far-off resonance from the ground and the biexciton levels, it is valid for the initial evolution, when the intracavity photon number does not saturate the QD transitions, and takes the form \begin{eqnarray}
   \label{rateEq1}
   \dot p_{n}&=&-\Gamma_n^{\rm out} p_{n}\nonumber\\
   &&+G_{n-1}^{(1)}p_{n-1}+G_{n-2}^{(2)}p_{n-2} \nonumber\\
   &&+\Gamma^{(1)}_{n+1} p_{n+1}+\Gamma^{(2)}_{n+2}p_{n+2}\,,
\end{eqnarray}
where $p_n$ denotes the probability that the cavity mode contains $n$ photons. The details of the derivation are given in Appendix~\ref{app:rate}. The rates take the form
\begin{eqnarray}
&&G^{(1)}_{n}=(n+1)A_1\,,\\
&&G^{(2)}_{n}=(n+1)(n+2)A_2\,,\\
&&\Gamma^{(1)}_{n}=n C_1\,,\\
&&\Gamma^{(2)}_{n}=n(n-1)C_2\,,\\
&&\Gamma_n^{\rm out}=G^{(1)}_{n}+G^{(2)}_{n}+\Gamma^{(1)}_{n}+\Gamma^{(2)}_{n}\,,
\end{eqnarray}
and they describe transitions connecting different photonic states, as shown in Fig.~\ref{figRate1}. The coefficients appearing in the expressions are rates. In particular, $A_1$ and $A_2$ are the one- and two-photon gain, respectively,
\begin{eqnarray}
A_1&=&\frac{8g^2P\pq{P(P+3\Gamma_d)+\gamma(6P+\Gamma_d)+\gamma^2}}{\pt{P+\gamma}^2\chi^2}\,,\nonumber\\
A_2&=&\frac{32g^4P^2}{\pt{P+\gamma}^2(P+\gamma+\Gamma_d)\chi^2}\,,
\end{eqnarray}
while $C_1$ and $C_2$ are rates of dissipation of one and two photons, respectively, and read
\begin{eqnarray}
C_1&=&\kappa+\frac{\gamma}{P}A_1\,,\nonumber\\
C_2&=&\frac{\gamma}{P}A_2\,.
\end{eqnarray}
Equation~(\ref{rateEq1}) has the form of the standard rate equation for lasing~\cite{scully_book02a}, where in addition it includes two-photon processes. Differing from previous treatments on two-photon lasing, based on injecting atoms into a resonator prepared in the upper states, rate equation~(\ref{rateEq1}) is derived for a medium which is continuously pumped inside the resonator. {This approximation is adequate when the biexciton transition is not saturated, i.e. for low excitation of the system}. In particular, the coefficients of the one-photon processes correspond to the coefficients of the laser rate equation  (see, for instance, Ref.~\onlinecite{scully_book02a}) here expanded to lowest order in the field-medium coupling.

Comparison between the different rates provides insight in the initial dynamics of the cavity field. In particular, gain overcomes dissipation when the pumping rate exceeds the losses, $P>\gamma,\kappa$ (we recall that this rate equation has been obtained in the limit of small $\kappa$, see Appendix~\ref{app:rate}). Moreover, two-photon gain will be (initially) dominant over one-photon gain when the ratio
\begin{equation}
\frac{G^{(2)}_n}{G^{(1)}_n}=\frac{(n+2)A_2}{A_1}>1\,.
\end{equation}
For $n=0$, this reduces to the inequality
\begin{equation}
\label{gain}
1<\frac{2A_2}{A_1}\sim \frac{8 g^2}{P^2}
\end{equation}
which has been obtained in the limit $P\gg\gamma,\Gamma_d$, showing that two-photon gain overcomes one-photon gain provided that the cavity-exciton coupling strength is larger than the pumping rate. We remark that condition~(\ref{gain}) has been obtained assuming a very good resonator, in the limit in which the cavity loss rate is the smallest parameter of the dynamics. We further note that condition~(\ref{gain}) is consistent with the semi-classical result in Ref.~\cite{ning04a} obtained for a quantum well, when studying two-photon amplification of a weak classical probe. Comparison with the numerical results at large cooperativity, however, shows that the signatures of two-photon lasing are found when the pumping rate exceeds the coupling strength, contrarily to condition~(\ref{gain}). We attribute this discrepancy to the fact that lasing in a high-$Q$ cavity is thresholdless~\cite{rice94}, therefore, the comparison of the two- and one-photon gain at the initial stage of the dynamics does not provide sufficient information about the behavior at steady state.

\section{Discussion and conclusions}
\label{sec:Conclusions}

We have analyzed the conditions for two-photon gain and lasing with a single quantum dot inside a high-$Q$ microcavity. We have shown that, by taking advantage of the biexciton binding energy, the system can exhibit clear features of a two-photon emitter when the cavity frequency is tuned to the two-photon resonance. Two-photon lasing, in particular, is found in high-$Q$ cavity for sufficiently large pump rates.

A crucial point to be stressed is the feasibility of two-photon lasing by using QD embedded in semiconductor microcavities. Different techniques have been developed for fabricating systems in which just a single QD is located close to the maximum (antinode) of the electromagnetic mode of a cavity built up either in a photonic crystal \cite{winger08,badolato05,hennessy07a,nomura09} or in a  micropillar \cite{ates09a,ates09b,dousse09}. In order to establish whether or not we have been working within a reasonable regime of parameters, we will compare our values with the ones used in two of the most recent experiments \cite{winger08,nomura09} with incoherent pumping as it is the case of our theory. One of the strongest coupling reported up to now~\cite{winger08} is $g=150\mu$eV in a system with $\kappa = 83\mu{\rm eV} =0.55g$ and $\delta$ in the range from $100$ to $200\mu$eV i.e. $\delta \approx g$. Smaller but similar values are those of Ref.~\cite{nomura09} where $g=68\mu$eV, $\kappa = 39\mu{\rm  eV} =0.57g$.
These sets of parameters are not far from the ones we use for most of the results we present here which have been obtained for a ratio $\kappa/g =0.1$ which corresponds to $17\%-18\%$ of that achieved in the experiments. Devices with such characteristics are expected to be realizable in the near future~\cite{auffeves}.

Taking $\gamma < \kappa$ (see figures \ref{fig:ssQD} or \ref{spectrum2}) is also very reasonable compared with currently existing systems. { A} crucial magnitude for getting two photon lasing { seems to be} the biexciton binding energy $\chi$ that we typically take as 20 times $g$. With the experimental values above mentioned for $g$ this means $\chi $ between $1.4$meV and 3~meV, which are very typical values for self-organized growth QD \cite{skolnick04}.

The comparison of the pumping rate $P$ with experimental values is more complicated. One must consider two { issues}: the evaluation of the externally applied power excitation and its reduction to the absorbed pumping. The first factor, the externally applied pump $P_{ext}$, is just the number of photons arising to the sample per unit time. This can be evaluated as the pumped energy per unit time, i.e. the pump power, over the energy of one of those photons. In the case of Ref.~\cite{winger08} a pump power of 1~nW and a wavelength of $818 nm$ gives $P_{ext}=0.041 {\rm ps}^{-1}$ which in units of energy becomes $P_{ext} =170\mu$eV. In~\cite{nomura09} the wavelength is $800$~nm while pump power is varied from 25~nW to 500~nW, which gives $418\mu{\rm eV} \leq P_{ext} \leq 10474\mu {\rm eV}$. In the second step, getting $P$ from $P_{ext}$ requires to estimate the reduction due to the reflection of the light by the sample, the absorption losses in mirrors or the part of the light which is actually absorbed by the cavity itself. This reduction from external to absorbed pumping must be evaluated for each particular case but a reasonable estimation for the reduction is to get $P$ between $1\%$ and $5\%$ of $P_{ext}$ \cite{rohner97,strauf06}. This means that in the case of reference \cite{winger08} one can estimate $P \approx 4.5 \mu {\rm eV} = 0.03g$ while in the case of reference \cite{nomura09} $12\mu {\rm eV} \leq P \leq 300 \mu {\rm eV}$ which means $0.18g \leq P \leq 4.5g$. The results we present in this paper are obtained with $0.01 g \leq P \leq 5 g$ which are values precisely in the regime of the experiments discussed above.

The conclusion from the previous analysis shows that the two photon lasing, as depicted by our results, can be feasible with currently available samples and technology for QD embedded in microcavities.

\begin{acknowledgments}
We thank Atac {\u Imamo\=glu} and Sarah Kajari-Schr\"oder for helpful comments. This work has
been supported by the European Commission (EMALI, MRTN-CT-2006-035369; Integrated Project SCALA, Contract No.\ 015714), by the European Science Foundation (EUROQUAM ``CMMC''), and by the Spanish MEC under contracts QOIT Consolider-Ingenio 2010 CSD2006-0019; QNLP, FIS2007-66944 and MAT2008-01555/NAN, and by CAM under contract S-0505/ESP-0200.  E.d.V. acknowledges support of a Newton Fellowship. A.G.T. acknowledges support from \emph{La Caixa}. G.M. is supported by the DFG through a Heisenberg-Professorship. The Ramon-y-Cajal, Juan-de-la-Cierva, and FPU programs of the Spanish Ministry of Education and Science (MEC) are gratefully acknowledged.
\end{acknowledgments}

\appendix

\section{Effective Dynamics and rate equation}
\label{app:rate}

In this appendix, we first derive the effective master equation for QD and cavity mode in the regime in which $\chi$ is the largest parameter. We start from the master equation~(\ref{Lincoh}), and use the basis of symmetric and antisymmetric superposition of the exciton states,
\begin{eqnarray}
    \ke{\pm }=\frac{1}{\sqrt{2}}\pt{\ke{E_1}\pm\ke{E_2}},
\end{eqnarray}
noting that the cavity mode couples with the symmetric state. In this basis Hamiltonian~(\ref{eq:1:0}) takes the form
\begin{eqnarray}
  H&=&\frac{\chi+\delta}{2}\pt{\kb{+}{+}+\kb{-}{-}}+\delta \kb{B}{B}\nonumber\\
  &&+\frac{\epsilon}{2}\pt{\kb{+}{-}+\kb{-}{+}}\nonumber\\
  &&+ \sqrt{2}g \pq{\pt{\kb{G}{+}+\kb{+}{B}}a\da  + {\rm H.c.}}\,,
\end{eqnarray}
where $\epsilon=\omega_1-\omega_2$ is the frequency difference between the exciton states, which we assume $|\epsilon|\ll\chi$. In the limit in which $\chi$ is the largest parameter, the single-photon coherences $\rho_{j,G}(n,m)=\br{j,n}\rho\ke{G,m}$, and $\rho_{B,j}(n,m)=\br{B,n}\rho\ke{j,m}$ (with $m=n,n+ 1$) oscillate very rapidly with respect to the other elements of the density matrix, and can be adiabatically eliminated from the dynamics, obtaining the set of equations
\begin{eqnarray}
\label{eqsrho}
\dot\rho_{GG}&=&-2P\rho_{GG}+\gamma(\rho_{++}+\rho_{--})\nonumber\\&&
+\lpq{-\pt{\beta+{\rm i}g_{\rm eff}}
\pt{a\da a\rho_{GG}+{a\da}^2\rho_{BG}-a\da\rho_{++}a}  }\nonumber\\&&\rpq{
+{\rm H.c.} }+\frac{\kappa}{2}{\cal L}(a)\rho_{GG}\,,\nonumber\\
\dot\rho_{BB}&=&-2\gamma\rho_{BB}+P(\rho_{++}+\rho_{--})\nonumber\\&&
+\lpq{-\pt{\alpha+{\rm i}g_{\rm eff}}
\pt{aa\da\rho_{BB}+a^2\rho_{GB}-a\rho_{++}a\da}   }\nonumber\\&&\rpq{+{\rm H.c.} }+\frac{\kappa}{2}{\cal L}(a)\rho_{BB}\,,\nonumber\\
\dot\rho_{GB}&=&-(P+\gamma+\Gamma_d-{\rm i}\delta)\rho_{GB}\nonumber\\&&
-\pt{\alpha+{\rm i}g_{\rm eff}}
\pt{a\da a\rho_{GB}+{a\da}^2\rho_{BB}-a\da\rho_{++}a\da}\nonumber\\&&
-\pt{\beta-{\rm i}g_{\rm eff}}
\pt{\rho_{GB}aa\da +\rho_{GG}{a\da}^2-a\da\rho_{++}a\da}\nonumber\\&&
+\frac{\kappa}{2}{\cal L}(a)\rho_{GB}\,,\nonumber\\
\dot\rho_{BG}&=& {\dot\rho_{GB}}\da\,,\nonumber\\
\dot\rho_{--}&=&-\pt{P+\gamma+\frac{\Gamma_d}{2}}\rho_{--}
+{\rm i}\frac{\epsilon}{2}(\rho_{-+}-\rho_{+-})\nonumber\\&&
+P\rho_{GG}+\gamma\rho_{BB} +\frac{\Gamma_d}{2}\rho_{++}+\frac{\kappa}{2}{\cal L}(a)\rho_{--}\,,
\nonumber\\
\dot\rho_{++}&=&-\pt{P+\gamma+\frac{\Gamma_d}{2}}\rho_{++}
-{\rm i}\frac{\epsilon}{2}(\rho_{-+}-\rho_{+-})\nonumber\\&&
+P\rho_{GG}+\gamma\rho_{BB} +\frac{\Gamma_d}{2}\rho_{--}\nonumber\\&&
+\lpq{-\pt{\beta-{\rm i}g_{\rm eff}}
\pt{aa\da \rho_{++}-a\rho_{GB}a-a\rho_{GG}a\da} }\nonumber\\&&
-\pt{\alpha-{\rm i}g_{\rm eff}}
\pt{a\da a\rho_{++}-a\da\rho_{BG}a\da-a\da\rho_{BB}a}   \nonumber\\&&\rpq{+{\rm H.c.}}+\frac{\kappa}{2}{\cal L}(a)\rho_{++}\,,\nonumber\\
\dot\rho_{-+}&=&-\pt{P+\gamma+\frac{\Gamma_d}{2}}\rho_{-+}+\frac{\Gamma_d}{2}\rho_{+-}\nonumber\\&&
+2\pt{\beta+{\rm i}g_{\rm eff}}\rho_{-+}aa\da
+2\pt{\alpha+{\rm i}g_{\rm eff}}\rho_{-+}a\da a\nonumber\\&&
+{\rm i}\frac{\epsilon}{2}(\rho_{--}-\rho_{++})+\frac{\kappa}{2}{\cal L}(a)\rho_{-+}\,,\nonumber\\
\dot\rho_{+-}&=& {\dot\rho_{-+}}\da\,,
\end{eqnarray}
where $\rho_{ij}$ is here an operator over the Hilbert space of the cavity mode, and
\begin{eqnarray}
g_{\rm eff}&=&-4\frac{g^2}{\chi}\,,\\
\alpha&=&4\frac{g^2\pt{P+3\gamma+3\Gamma_d}}{\chi^2}\,,\\
\beta&=&4\frac{g^2\pt{3P+\gamma+\Gamma_d}}{\chi^2}\,.
\end{eqnarray}
\begin{figure}[!th]
  \begin{center}
  \includegraphics[width=6cm]{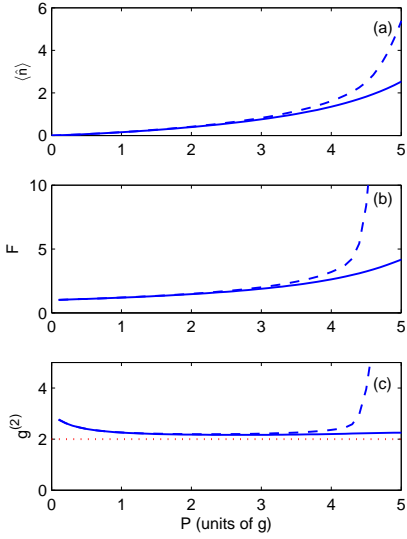}
    \caption{(Color online) Average cavity photon number $\mean{\hat n}$, Fano factor $F$ and $g^2(0)$, evaluated numerically from Master equation~(\ref{Lincoh}) (solid lines) and analytically using rate Eq.~(\ref{rateEq1}) (dashed lines) as a function of the pumping rate $P$ (in units of $g$). The parameters are $\chi=50g$, $\delta=\delta_{2P}$, $\Gamma_d=0$, $\gamma=4g$ and $\kappa=0.02g$. When the pump is sufficiently small, $P<\gamma$, such that the number of intracavity photon does not saturate the biexciton-ground state transition, the rate equation provides a good description of the dynamics at all times. In this regime the system behaves as a laser below threshold, and the cavity field approaches a thermal state. }
    \label{fig2}
  \end{center}
\end{figure}

The dynamics these equations describe is valid up to second order in the expansion in $1/|\chi|$. Moreover, we assume that the cavity decay rate $\kappa$ and the frequency difference $|\epsilon|$ are the smallest parameters, such that $\kappa,|\epsilon|\lesssim g_{\rm eff}$. Hence, the regime of validity of these equations is quite small and experimentally demanding. Nevertheless, they lead us to an analytic solution, which give us some insight into the process of two-photon lasing.

In the following we assume $\epsilon=0$, so that the last two equations in Eqs.~(\ref{eqsrho}) are decoupled from the rest, and the system dynamics can be described by the first six equations of  Eqs.~(\ref{eqsrho}). We first note that the coherent dynamics described by this set of equation, obtained by setting $P=\gamma=\Gamma_d=\kappa=0$, is equivalent to the two-photon dynamics described by the effective Hamiltonian derived in Eq.~(\ref{Heff}). When considering the incoherent processes, on the other hand, one finds that the exciton states are involved in the system dynamics in a nontrivial way.

We now aim at getting further insight into the regime, in which gain can be observed, and in particular in which two-photon gain is dominant. At this purpose we derive from Eqs.~(\ref{eqsrho}) a rate equation for the photonic state occupations. We assume $P+\gamma\gg |g_{\rm eff}|.$ In this limit the atomic population and the two-photon coherences decay very fast with respect to the typical time scale of the evolution of the cavity field. Denoting by
\begin{eqnarray}
    p_n={\rm Tr}\{\ke{n}\br{n}\rho\}
\end{eqnarray}
the population of the photon state $|n\rangle$, one obtains the rate equations for the population $p_n$, reported in Eq.~(\ref{rateEq1}) by expanding Eqs.~(\ref{eqsrho}) at lowest relevant order in $g_{\rm eff}/(P+\gamma)$ and taking the corresponding trace.

We remark that rate equation~(\ref{rateEq1}) has been obtained assuming very small cavity decay (high-$Q$ resonators) by an expansion in the QD-cavity field interaction strength, and therefore it does not contain the higher-order, nonlinear terms which describe the lasing regime. It is valid when these terms are negligible, namely, either (i) at all times when the pump is small such that the system behaves as a laser below threshold, in which the field is thermal (see Fig.~\ref{fig2}), or (ii) at the initial times, when the number of cavity photons is still sufficiently small that the nonlinear terms are negligible.

\section{Validity of the model}

\begin{figure*}[!th]
  \begin{center}
  \includegraphics[width=15cm]{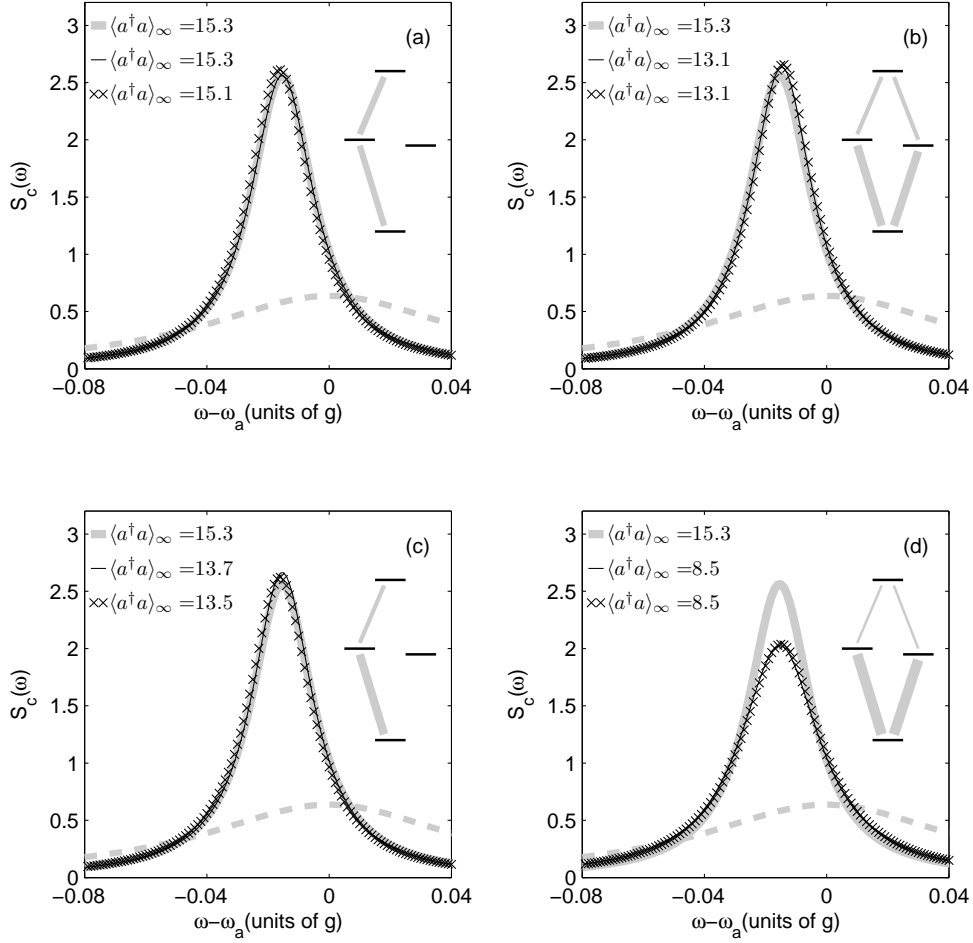}
    \caption{Cavity emission spectrum. The solid-thick-gray lines correspond to the situation, in which all four transitions have equal coupling strength, namely, $\alpha=\beta=1$, $g_{G1}=g_{1B}\equiv g$, while the dashed-thick-gray lines give the cavity emission spectrum in absence of coupling with the quantum dot. The solid black lines are found for (a) $\alpha=\beta=0$, $g_{G1}=g_{1B}=\sqrt{2}g$, $\delta=-0.2g$, $\epsilon=0$; (b) $\alpha=\beta=1$, $g_{G1}=1.5g$, $g_{1B}=0.66g$, $\delta=-3g$, $\epsilon=0$; (c) $\alpha=\beta=0$, $g_{G1}=2g$, $g_{1B}=g$, $\delta=-2.7g$, $\epsilon=0$; (d) $\alpha=\beta=1$, $g_{G1}=2g$, $g_{1B}=0.5g$, $\delta=-4.2g$, $\epsilon=0$. The crosses are found by setting $\epsilon=g$. The other parameters are $\chi=20g$, $\gamma=0.05g$, $\Gamma_d=0$, $P=1.8g$, $k=0.1g$. The effective two-photon coupling is the same for all the curves: $\tilde g_{\rm eff}=-g^2/\chi$.  In each plot it is explicitly indicated the steady state average photon number corresponding to each curve. The insets depict the coupling scheme for each set of parameters.
 }
    \label{fig15}
  \end{center}
\end{figure*}

In this appendix we study the limit of validity of the model introduced in section~\ref{Sec:Model}. In particular we focus on the simplifying assumptions of  equal coupling constants between the cavity field and the four electronic transitions and on the assumption of negligible detuning between the single exciton states.

In general the exciton states energies are not degenerate. We denote their detuning by $\epsilon$. Moreover, we denote the coupling strengths between the cavity mode and the four electronic transitions by the symbols $g_{G1}$, $g_{G2}$, $g_{1B}$ and $g_{2B}$.  The Hamiltonian, describing the coupling between QD and cavity field, takes the form
\begin{eqnarray}
\tilde H&=&\frac{\hbar}{2}\pq{\pt{\chi+\delta+\epsilon}\kb{E_1}{E_1}+\pt{\chi+\delta-\epsilon}\kb{E_2}{E_2}}\nonumber\\
&&+\hbar\delta\kb{B}{B}\nonumber\\
&&+\hbar g_{G1}\pq{\pt{\kb{G}{E_1}+\alpha\kb{G}{E_2}}a\da+{\rm H.c.}}   \nonumber\\
&&+\hbar g_{1B}\pq{\pt{\kb{E_1}{B}+\beta\kb{E_2}{B}}a\da+{\rm H.c.}},
\end{eqnarray}
where we have introduced the dimensionless coefficients $\alpha$ and $\beta$, giving the ratios $\alpha=g_{G2}/g_{G1}$ and $\beta=g_{2B}/g_{1B}$. Under the assumption of large $\chi$, we eliminate the intermediate, exciton states and obtain the effective Hamiltonian
\begin{eqnarray}\label{tildeH}
\tilde H_{\rm eff}&=&\hbar\pt{\delta+\tilde\delta_{2P}+\tilde\delta_{n}a\da a}\kb{B}{B}\nonumber\\
    &&+\tilde g_{\rm eff}\pt{a^2\kb{B}{G}+{a\da}^2\kb{G}{B}}
\end{eqnarray}
with
\begin{eqnarray}
    \tilde\delta_{2P}&=&\frac{2g_{1B}^2}{\chi}\pt{1+\beta^2}
    \nonumber\\
\tilde\delta_{n}&=&\frac{2}{\chi}\pq{g_{G1}^2\pt{1+\alpha^2}-g_{1B}^2\pt{1+\beta^2}   }\nonumber\\
    \tilde g_{\rm eff}&=&-\frac{2 g_{G1}g_{1B}}{\chi}
    \pt{1+\alpha\beta}.
\end{eqnarray}
Differing from the ideal case discussed in this article, where we take all coupling strength to be equal, we observe that in general the light shift of the two photon transition depends on the number of photons. However as far as the condition $\tilde\delta_{2P}+\tilde\delta_{n}\av{a\da a}\ll\chi$ is fulfilled, so that the perturbative expansion is valid, then the detuning $\delta$ can be set in order to minimize the effect of the light shift. In this case equal values of the effective coupling $\tilde g_{\rm eff}$ give similar dynamics independently of the exact values of the coupling constants $g_{G1}$, $g_{G2}$, $g_{1B}$ and $g_{2B}$. Moreover, the finite value of $\epsilon$ affects only the single exciton dynamics (see also Eq.~\ref{eqsrho}), but it is not relevant for the two photon transition, provided that the perturbative treatment, allowing for eliminating the intermediate levels, is valid.

In Figure~\ref{fig15} we report the cavity emission spectrum for various cases. In (a) we compare the cavity emission spectrum obtained with the simplified model of Sec.~\ref{Sec:Model} with the spectrum found when only one exciton state is coupled to the cavity; in (b) we study the situation in which the coupling with the ground-exciton transitions is larger than that with the exciton-biexciton transitions; in (c) only one exciton state is coupled to the cavity and the couplings to the ground and biexciton states are different; finally (d) is found by increasing the difference between the coupling with the ground-exciton and the exciton-biexciton transitions. In all the cases we consider both the situation in which $\epsilon=0$ and $\epsilon\neq 0$; moreover the parameters are set in order to keep $\tilde g_{\rm eff}$ equal for all the curves. The detuning $\delta$ for each curve is found by maximizing the  average steady-state cavity photon number as a function of $\delta$ for the corresponding set of parameters.
We observe that similar results for the cavity field are found in Fig.~\ref{fig15} (a), (b) and  (c). On the other hand, a relevant difference is found in (d) when the couplings with ground-exciton and exciton-biexciton transition are sufficiently separate in frequency.

In conclusion, similar two photon laser dynamics can be obtained, independently of the exact values of the coupling constants and exciton detuning, as far as the parameters that scales the effective two-photon dynamics are similar.
Analogous considerations are valid also for the effect of the spontaneous emission rates which are in general different for the four electronic transition: the main features of the laser dynamics are independent from the exact values of the individual rates.


\end{document}